\documentclass[aip,amsmath,amssymb,reprint]{revtex4-1}

\usepackage{graphicx}% Include figure files
\usepackage{dcolumn}% Align table columns on decimal point
\usepackage{bm}% bold math
\usepackage[utf8]{inputenc}
\usepackage[T1]{fontenc}
\usepackage{mathptmx}
\usepackage{etoolbox}
\usepackage{color}
\usepackage{bm}
\usepackage{bbm,mathbbol}
\usepackage{braket,bigints}
\usepackage{stmaryrd,mathtools}
\newtheorem{lemma}{Lemma}
\def \beq {\begin{equation}}
\def \eeq {\end{equation}}

\newcommand\independent{\protect\mathpalette{\protect\independenT}{\perp}}
\def\independenT#1#2{\mathrel{\rlap{$#1#2$}\mkern2mu{#1#2}}}
\makeatletter
\def\@email#1#2{%
 \endgroup
 \patchcmd{\titleblock@produce}
  {\frontmatter@RRAPformat}
  {\frontmatter@RRAPformat{\produce@RRAP{*#1\href{mailto:#2}{#2}}}\frontmatter@RRAPformat}
  {}{}
}%
\makeatother
\begin{document}

\preprint{AIP/123-QED}

\title{Quantum advantage in biometric authentication with single photons}
% Force line breaks with \\
\author{I. K. Kominis}
\affiliation{Department of Physics, University of Crete, Heraklion 71003, Greece}
\affiliation{Institute of Theoretical and Computational Physics, University of Crete, Heraklion 70013, Greece}
\author {M. Loulakis}
\affiliation{School of Applied Mathematical and Physical Sciences, National Technical University of Athens, 15780 Athens, Greece}
\affiliation{Institute of Applied and Computational Mathematics, FORTH, 71110 Heraklion, Greece}
\date{\today}

\begin{abstract}
It was recently proposed to use the human visual system's ability to perform efficient photon counting in order to devise a new biometric methodology. The relevant biometric "fingerprint" is represented by the optical losses light suffers along several different paths from the cornea to the retina. The "fingerprint" is accessed by interrogating a subject on perceiving or not weak light flashes, containing few tens of photons, so that the subject's visual system works at the threshold of perception, at which regime optical losses play a significant role. Here we show that if instead of weak laser light pulses we use quantum light sources, in particular single-photon sources, we obtain a quantum advantage, which translates into a reduction of the interrogation time required to achieve a desired performance. Besides the particular application on biometrics, our work further demonstrates that quantum light sources can provide deeper insights when studying human vision. 
\end{abstract}

\maketitle

\section{Introduction}
Secure biometric identification \cite{Unar,Yang} has some analogies to secure transmission of information: in both cases the potential for some impostor to maliciously intervene in the respective process must be avoided. Quantum cryptography \cite{Gisin,Pirandola} has served as a paradigm of the so-called "quantum advantage", i.e. the security cryptographic key transmission is guaranteed by the laws of quantum physics. In contrast, classical means of information transmission are vulnerable to e.g. eavesdropping. 

In a similar fashion, we recently proposed \cite{Loulakis} a biometric authentication scheme which takes advantage of the human visual system's ability to perform photon counting. There, the laws of photon statistics and photodetection at the quantum level of a small number of photons are central for the workings of the scheme, providing for an uncompromising security against an impersonator. The relevant "fingerprint" is a physical property of the visual system, including the eyeball, retina and brain. In other words, the biometric authentication process rests on the conscious perception of weak-intensity light, working at the very threshold of visual perception of a specially designed light stimulus. In particular, the "fingerprint" in this method is the optical loss suffered by light along its propagation from the cornea to the retina. 

In our previous work \cite{Loulakis} the visual stimulus was supposed to originate from laser light. However, the number of photons in a pulse of laser light is known to fluctuate from pulse to pulse around a mean value, and these Poissonian fluctuations limit the performance of the authentication process as will be outlined later. Interestingly, quantum optical technology can now offer "quantum light" sources, in particular single photon sources, producing light with much narrower distribution of the photon number. We will here show how such a quantum light source translates into a concrete quantum advantage of the authentication process, namely a reduction in the interrogation time. 

Besides its applied perspective, this work further signifies the promise of the emerging field of quantum vision \cite{qv1,qv2,qv3,qv4,qv5,qv6}, one aspect of which is the use of quantum light sources to study human or animal vision at a deeper level than was previously possible with classical light sources.
\section{Preliminaries}
Our biometric authentication methodology was inspired by the early experiment of Hecht et al. \cite{Hecht}, eloquently described in modern terms by Bialek \cite{Bialek}. Hecht et al. unambiguously demonstrated that rod cells, the scotopic photoreceptors in our retina, are efficient photon detectors, and moreover, they identified a threshold in the number of detected photons for visual perception to take place. We denote this threshold by $K$, found \cite{Hecht} to be $K\approx 6$. Parenthetically, a recent psychophysical experiment \cite{Vaziri} performed along similar lines, but using a modern single-photon source for the stimulus light, found that $K\approx 1$. In any case, to our understanding, the precise value of $K$ and its possible relation to the particular physiological condition of the subject is still an open problem.

In more detail, the three authors in \cite{Hecht} had their eyes illuminated by very weak-intensity light pulses, with the integrated photon number within each pulse being so small, that the visual perception became probabilistic, with the probability of seeing denoted by $P_{\rm see}$. An expression for $P_{\rm see}$ can be found as follows. Denote by $\tilde{N}$ the mean number of photons within a light pulse of duration $\tau$. Coherent light has Poissonian photon statistics, i.e. the probability to have exactly $n$ photons within such a pulse is $\tilde{N}^ne^{-\tilde{N}}/n!$. Of course the authors in \cite{Hecht} did not have lasers, however it is known that for averaging times (in this case the light pulse duration $\tau$) longer than the correlation time of a classical light source the photon statistics are again Poissonian \cite{Fox}.

However, when the mean number of photons per pulse incident on the eyeball is $\tilde{N}$, the actual mean number of photons per pulse {\it detected} by the retina's photoreceptors is reduced by a factor $0<\alpha<1$. This factor describes the optical losses suffered by light along its path from the cornea to the retina, as well as the probability of photodetection by the illuminated photoreceptors. Now, the probability that the number of photons detected by the illuminated patch of the retina is exactly $n$ is given by $(\alpha\tilde{N})^ne^{-\alpha\tilde{N}}/n!$. If this number is higher than the detection threshold $K$, whatever its value, then the perception of "seeing" a spot of light will take place. Thus, the probability of seeing is 
\begin{equation}
P_{\rm see}=\sum_{n=K}^{\infty}{{(\alpha\tilde{N})^ne^{-\alpha\tilde{N}}}\over{n!}}
\end{equation}
\begin{figure*}
\begin{center}
\includegraphics[width=16.5 cm]{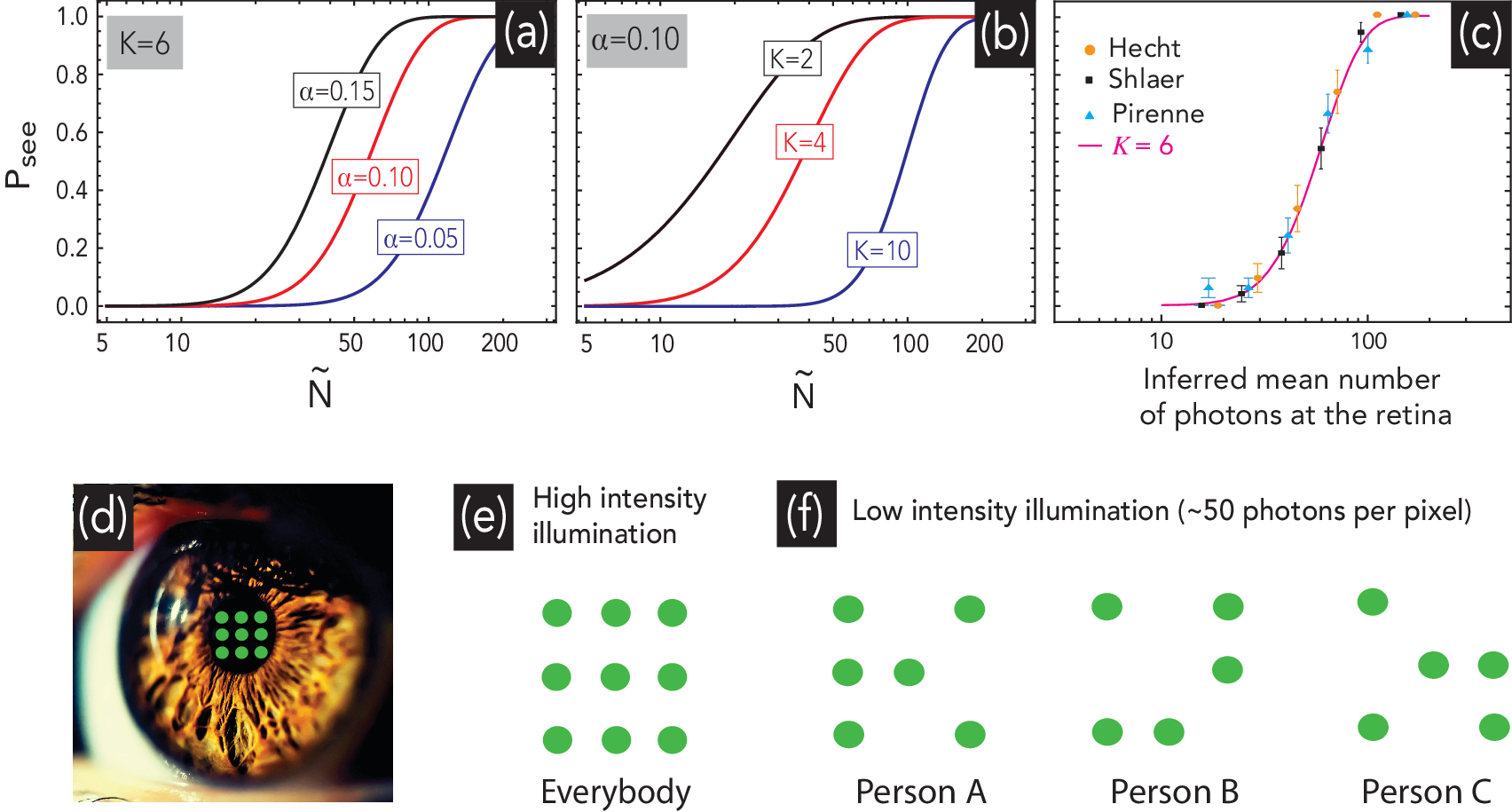}
\caption{Probability of seeing a light pulse having mean incident photon number per pulse $\tilde{N}$ versus $\tilde{N}$, as calculated from Eq. 1, for (a) various values of the optical loss parameter $\alpha$, and constant perception threshold $K$, and (b) various values of $K$ and constant $\alpha$. (c) Experimental results of \cite{Hecht} for the probability of seeing by the three authors (figure adapted from \cite{Bialek}). In this figure the x-axis is the inferred number of photons at the retina, i.e. the authors \cite {Hecht} have shifted the curves of $P_{\rm see}$ along the x-axis using some factor $\alpha$ for each one of the three authors. This results in a "universal" response apparently having a common threshold $K$. For these authors, the difference in the $\alpha$ value among them was a nuisance for what they were trying to achieve, i.e. demonstrate the single photon detection capability of rod cells. For us, this difference in $\alpha$ is taken advantage of to define our biometric quantifier. (d-f) Simplified presentation of the idea behind the biometric authentication using the photon counting capability of the human visual system. (d) A light stimulus source is supposed to provide for parallel laser beams patterned in an array, here shown as a $3\times 3$ array. The laser beams propagate in parallel from the source to the eye, being incident on the cornea, all being simultaneously illuminated during a given pulse. If one could see the reflection of the laser beams off the cornea, one would see the image (d), where the pupil is shown to be illuminated by 9 spots. (e) With each laser beam containing a large number of photons per pulse, and further assuming that the human subject being illuminated is myopic, everybody would report seeing 9 different spots patterned in such a $3\times 3$ array. (f) However, if the number of photons per beam per pulse is reduced to the regime of 10-100 photons, the visual perception would be working close to its threshold. In such a case, the optical losses suffered by light along these 9 different paths, different among paths for each individual, and different for a "geometrically similar" path among individuals, will result in a different perception pattern for each subject.}
\label{fig1}
\end{center}
\end{figure*}

In Fig. \ref{fig1} we plot examples of the dependence of the probability $P_{\rm see}$ on the mean number of photons per pulse incident on the cornea, $\tilde{N}$. In Fig. \ref{fig1}a we keep the threshold $K$ constant at the value of $K=6$, and change the loss factor $\alpha$, whereas in Fig. \ref{fig1}b we keep $\alpha$ constant at the value $\alpha=0.10$, and change $K$. Both dependences are rather obvious to interpret. It is important to note that the change of $\alpha$ (Fig. \ref{fig1}a) hardly changes the overall shape of the functional dependence of $P_{\rm see}$ versus $\tilde{N}$, and essentially shifts the figure along the x-axis. In contrast, the change of $K$ qualitatively changes the shape of the dependence of $P_{\rm see}$ versus $\tilde{N}$. Now, although each one of the three authors in \cite{Hecht} produced a different dependence of $P_{\rm see}$ versus $\tilde{N}$, all three curves could be coalesced by such a translation along the x-axis, and all could be fit with a common value of $K\approx 6$. This is shown in Fig. \ref{fig1}c.

The authors in \cite{Hecht} managed to make a remarkable case, despite the presence of a subjective observable, as is the optical loss parameter $\alpha$, which changes among individuals and perplexes the analysis of individuals' responses to perceiving or not faint light pulses. The case is about two objective properties of the human visual system. The first has to do with the wiring of the photoreceptor cells to deeper neural layers communicating visual responses to the brain. This wiring determines the perception threshold $K$, which appears to be a common systemic property. The second is that retina's photoreceptors are efficient single photon detectors. This follows from the fact that the experimentally inferred number of photons at the retina (see $x$-axis of Fig. \ref{fig1}c) is much smaller than the number of illuminated rod cells. It took several years until the quantum photo-detection properties of rod cells were unraveled with modern photonic and quantum-optical technology \cite{Baylor1,Baylor2,Baylor3,Baylor4,Kriv1,Kriv2,Kriv3,Nelson}.
\section{Quantum Biometrics}
Our proposal on quantum biometrics \cite{Loulakis} essentially turns the coin around: the variability of the parameter $\alpha$ among individuals was a nuisance for Hecht et al., who were aiming at the single-photon-detection capability of rod cells. We now know this physiological capability is indeed the case. Instead, we would like to use the variability of the parameter $\alpha$ as a biometric quantifier. However, just one number is not enough as a biometric "fingerprint". Hence the idea put forward in \cite{Loulakis} is that the relevant "fingerprint" is a whole map of $\alpha$ values, the so-called $\alpha$-map. This results from considering several paths of light towards the retina, illuminating the retina at several different spots. This point has to do with visual optics, and will be further elaborated upon elsewhere. The crux of the matter is illustrated in Figs. \ref{fig1}d-f. Suppose we have an array of e.g. 9 laser beams, patterned in a $3\times 3$ matrix (Fig. \ref{fig1}d). Further suppose that these beams are all illuminated simultaneously, and moreover, let us assume that the mean photon number per beam per pulse is very large, say $\gg 100$ photons. In such a scenario {\it every} subject (without any visual deficiency) will report seeing 9 spots (Figure \ref{fig1}e). This is because with certainty, everybody will perceive a pulse containing a large number of photons (far right in the curves of Figs. \ref{fig1}a-c, where $P_{\rm see}\approx 1$). However, as we reduce the mean photon number per laser beam per pulse, and move to the regime of the visual threshold described by the variable $P_{\rm see}$ in Figs. \ref{fig1}a-c, each individual will report different patterns of perception, as shown in Fig. \ref{fig1}f. This difference in perception in the regime of the visual threshold is exactly what our biometric identification scheme takes advantage of.

To describe the methodology in more detail, we first note that the prerequisite is that the $\alpha$-map of the subject that will need to be authenticated by the biometric device has been already measured and stored. This is like taking a subject's fingerprint and registering it in the relevant database. The process of registering the subject's "fingerprint" for the first time, as well as the relevant ageing effects will be addressed elsewhere.

When the subject wants to be authenticated, the biometric device must implement a measurement protocol, the result of which is either positive or negative. Thus, two central quantifiers of its performance are the false-negative and false-positive probability, denoted by $p_{fn}$ and $p_{fp}$, respectively. The former is the probability that a subject truly claiming to be who he or she is, is {\it not} authenticated as such. The latter is the probability that an impostor, falsely claiming to be somebody else is positively identified as that other person. Obviously, the longer the authentication process, the smaller these two probabilities should become. Hence, a third important performance quantifier is how much time is required to achieve a given desired value for $p_{fn}$ and $p_{fp}$. 

To proceed, we note that central to the authentication algorithms described in \cite{Loulakis}, as well as the one described here is that we choose to work with retinal spots associated with either too high or too low values of $\alpha$. Let us call Alice the subject who appears and wishes to be positively authenticated. Eve will be an impostor maliciously claiming to be Alice. We will suppose that Eve is not aware of Alice's $\alpha$-map. As a result, Eve does not know whether the device is illuminating a low-$\alpha$ or a high-$\alpha$ spot of Alice, and thus cannot tune her responses accordingly. The spots being illuminated are randomly chosen by the device, and as far as Eve is concerned, they could be of any kind. Moreover, the device illuminates every spot, no matter of what kind, {\it with the same mean number of photons per pulse}. Thus, even if Eve is equipped with a perfect photon counter, she would just measure light pulses with a given mean number of photons. This measurement does not convey to her any useful information. Further, since she is not aware of Alice's $\alpha$-map, even if Eve is equipped with a perfect position-sensitive photon detector, she still cannot extract any useful information from any stimulus light patterns emitted by the biometric device. {\it Eve is forced to respond randomly to the device's interrogations on whether the subject does perceive or does not perceive the light flashes}. We will now quantify all of the above using a specific authentication protocol, which is a variant of the protocols given in \cite{Loulakis}. This variant, and its modification when using a quantum light source, are straightforward to understand.\subsection{Authentication protocol}
We assume that the device simultaneously illuminates $N$ different retinal spots, a random number $H$ of which are high-$\alpha$ spots, the rest being low-$\alpha$ spots. The interrogated subject is then questioned on how many spots she perceived. We accept her answer, $R$, as correct, if $R=H$. There are two competing mechanisms that push Alice towards failure. The possibility that she does not see an illuminated high-$\alpha$ spot, and the possibility that she does see an illuminated low-$\alpha$ spot. 
\subsubsection{Eve's strategy}
Suppose Eve comes along claiming to be Alice and asking to be authenticated. We assume that Eve has access to the identification protocol and knows the distribution of the random variable $H$. Further suppose that she is equipped with perfect position-sensitive photon counters. She thus finds that $N$ spots are being illuminated. Her answer on the device's interrogation about how many spots she perceived can be a random variable, call it $X$, which she chooses with some strategy. For sure, $X\independent H$, that is X is independent of $H$, since Eve does not know how many spots of which kind (high-$\alpha$ or low-$\alpha$) are being illuminated. Thus for her answer, she must use a random variable $X$ independent of $H$, chosen so that the probability $\mathbb{P}[X=H]$ is maximized. But it is
\begin{align}
\mathbb{P}[X=H]&=\sum_{m=0}^N\mathbb{P}[X=m,H=m]\nonumber\\
&=\sum_{m=0}^N\mathbb{P}[X=m]\mathbb{P}[H=m]
\end{align}
The second line follows from the independence of $X$ from $H$ and is a weighted average which is maximized when the random variable $X$ assumes those values $m$ which maximize the probability $\mathbb{P}[H=m]$, that is, the modes of the distribution of $H$. For example, if $H$ is unimodal, Eve's optimal response must be the mode of the distribution. In any case, since the range of values of $H$ is $\{0,1,...,N\}$, there must be some $m\in\{0,1,...,N\}$ for which $\mathbb{P}[H=m]\geq 1/(N+1)$.
Thus 
\begin{equation}
\max_{X\independent H} \mathbb{P}[X=H]\geq{1\over {N+1}}
\end{equation}
\subsubsection{Device's interrogation strategy}
In view of the preceding analysis, the biometric device should choose the distribution of $H$ to be as uninformative as possible, i.e. to be uniform on $\{0,1,...,N\}$. That is,
\[
\mathbb{P}[H=k]=\frac{1}{N+1},\qquad k=0,1,\ldots,N.
\] In such a case, regardless her answer, Eve will have probability ${1\over {N+1}}$ to answer correctly.
\begin{figure}
\begin{center}
\includegraphics[width=8.5 cm]{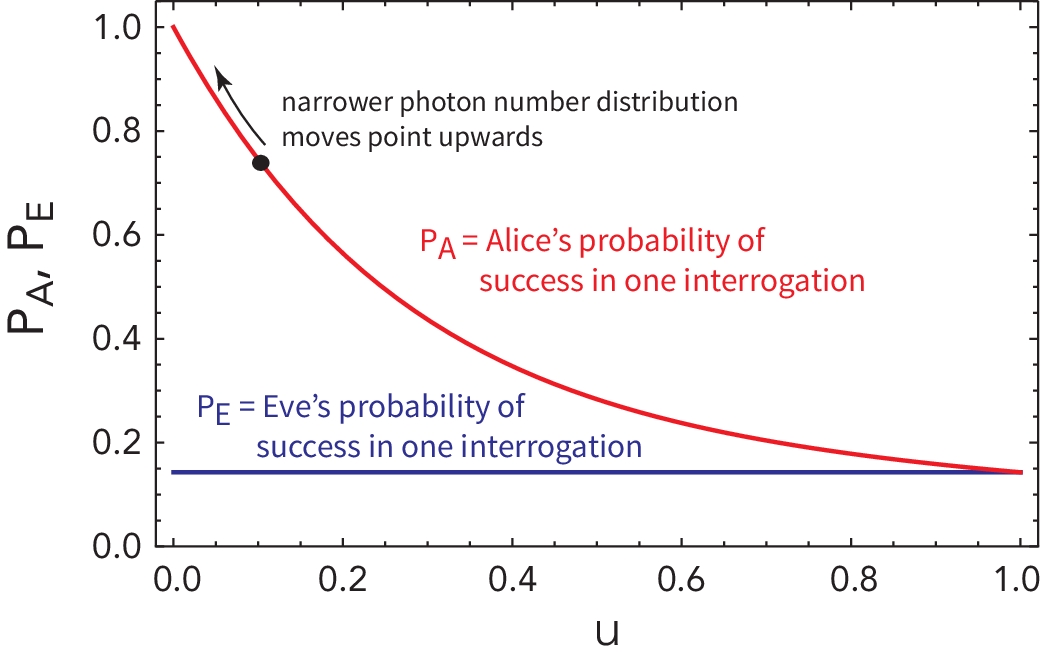}
\caption{Probability of Alice, $P_A$, and probability of Eve, $P_E$, correctly responding to one interrogation, consisting of illuminating a number of $N$ spots in total (some of which are high-$\alpha$ and the rest low-$\alpha$) and asking how many spots the subject perceived. The probability $P_A$ depends on $u=p_H+p_L$, which is the sum of the probability that Alice does not perceive an illuminated high-$\alpha$ spot, $p_H$, and the probability that  she does perceives a low-$\alpha$ spot, $p_L$. The value of $u$ depends on the photon statistics of the light, hence narrower photon number distributions suppress $u$ and thus provide for a quantum advantage. For this figure $N=6$.}
\label{PaPe}
\end{center}
\end{figure}
\subsubsection{Positive authentication of Alice}
We next consider Alice's probability of a successful answer, i.e. the probability $\mathbb{P}[Y=H]$, where $Y$ stands for the number of spots Alice perceives.
It is helpful to consider the number $E_{fn}$ of high-$\alpha$ spots that Alice fails to perceive, as well as the number $E_{fp}$ of low-$\alpha$ spots that Alice perceives.
Clearly, $Y= (H-E_{fn})+E_{fp}$, and Alice succeeds if the number of false-positive perceptions is equal to the number of false-negative perceptions.
If Alice fails to perceive a stimulus on a high-$\alpha$ spot with probability $p_H$, and she perceives a stimulus on a low-$\alpha$ spot with probability $p_L$, then, conditionally on $H=k,\ E_{fn}$  and $E_{fp}$ are independent binomial random variables, with parameters $(k,p_H )$ and $(N-k,p_L)$, respectively. The following Lemma is of interest on its own, and does not have an immediate combinatorial interprepation. The proof is given in the Appendix.
\begin{lemma}
Suppose $H$ is uniformly distributed on $[n]:=\{0,1,\ldots,N\}$ and, conditionally on $\{H=k\}$, the random variables $E_{fn}$ and $E_{fp}$ are independent binomial with parameters $(k,p)$ and $(N-k,q)$. Then,
\[
\mathbb{P}[E_{fn}=E_{fp}] = \frac{1-\big(1-(p+q)\big)^{N+1}}{(N+1)(p+q)}
\]
\end{lemma}
By setting $u=p_H+p_L$, a direct consequence of the Lemma is that  Alice's probability of success reads
\begin{equation}
\label{pAp}
P_A=P_A(N,u)=\frac{1-(1-u)^{N+1}}{(N+1)u},
\end{equation}
For example, for $N=6$ spots and $u=0.1$, Alice succeeds in one interrogation with probability $P_A\simeq 0.745$, while an impostor, Eve, succeeds with probability $P_E=1/7\simeq 0.143$. The probabilities $P_A$ and $P_E$ are plotted in Figure \ref{PaPe}. As will become clear in the following, the value of $u=p_H+p_L$ for Alice depends on the photon statistics of the illuminating light source, so this is the point where the quantum advantage comes in.
\begin{figure*}
\begin{center}
\includegraphics[width=17.8 cm]{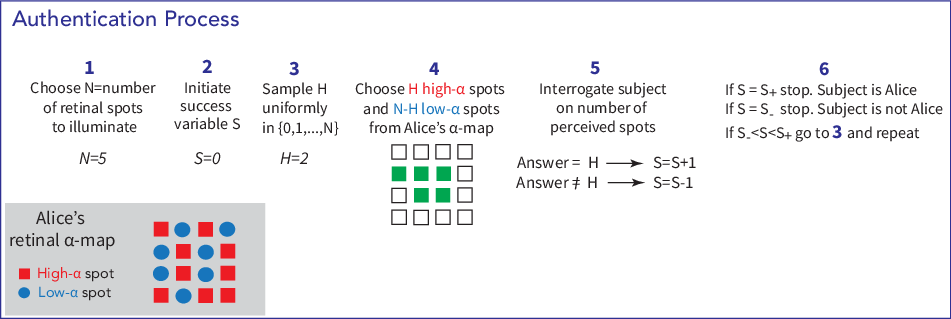}
\caption{Authentication algorithm. We suppose the biometric device has already measured and stored Alice's $\alpha$-map, and has classified her retinal spots (in this example in a grid of $5\times 5$ spots) into high-$\alpha$ and low-$\alpha$ spots, shown in the inset as red squares and blue circles, respectively. A subject approaches the device and claims to be Alice. The first step of the authentication algorithm is to choose a number $N$ of spots to illuminate, in this example $N=5$. Then, we initiate the success variable $S$ to the value $S=0$. In the  third step we uniformly sample $H\in\{0,1,\ldots,N\}$. In step 4, we choose $H$ high-$\alpha$ and $N-H$ low-$\alpha$ spots to illuminate. In this example $H=2$, which means that the device will illuminate 2 high-$\alpha$ spots and 3 low-$\alpha$ spots, as in step 4. In Step 5 the subject is interrogated on the number of spots she perceived, and if her answer is $H$, the success variable $S$ is updated by $S=S+1$, otherwise $S=S-1$. In the final step 6 the device compares $S$ with $S_-$ and $S_+$, which are defined in the text and are derived from the required performance on $p_{fp}$ and $p_{fn}$. As long as $S_-<S<S_+$, we repeat the algorithm going back to step 3. In case $S$ reaches $S_{+}$ the algorithm stops and the subject is authenticated as Alice, whereas if $S$ reaches $S_-$ the subject is not authenticated as Alice.}
\label{algorithm}
\end{center}
\end{figure*}
\subsubsection{Repeating the test to achieve a given $p_{fp}$ and $p_{fn}$}
In Figure \ref{algorithm} we show how the authentication algorithm works. As mentioned above, the biometric device has stored Alice's $\alpha$-map, so it can classify Alice's retinal spots into high-$\alpha$ spots, for which Alice has a high probability of seeing the light flash, and low-$\alpha$ spots, for which this probability is low. The device chooses the number $N$ of spots it will illuminate (step 1). In the example of Figure \ref{algorithm}, it is $N=5$. The specific value of $N$ is the result of an optimization to be presented in the following.

We also define a success parameter $S$, which is initiated at the value $S=0$ (step 2). Then, the device randomly selects a number $H\in\{0,1,\ldots,N\}$ (step 3), and illuminates $H$ spots of the type "high-$\alpha$" and $N-H$ spots of the type "low-$\alpha$" (step 4). In the example of Fig. \ref{algorithm}, it is $H=2$. The subject is then (step 5) interrogated on how many bright spots she perceived. If her answer equals $H$, then we update the success parameter to $S=S+1$, otherwise to $S=S-1$. In the last step of the algorithm $S$ is compared with a positive upper value $S_{+}>0$ and with a negative lower value $S_{-}<0$. If $S=S_{+}$ we stop the interrogation process, authenticating the subject as Alice. If $S=S_{-}$, then we stop the process with a negative authentication, i.e. the subject is not Alice. Otherwise, we return to step 3 of the algorithm.

The choice of $S_{+}$ and $S_{-}$ depends on the desired values of $p_{fp}$ and $p_{fn}$, as follows. Both Alice and Eve perform a random walk in $S$-space with unit step and probabilities $P_A$ and $P_E$, respectively, for a positive step, and $1-P_A$ and $1-P_E$, respectively, for a negative step. If by  $\tau_{S\pm}=\inf\{k\ge 0: S_k=S\pm\}$ we denote the interrogation round in which $S$ first reaches $S_{\pm}$, then \cite{Norris} the probability that Eve's success parameter reaches the value $S_{+}$ and thus Eve is falsely identified as Alice is
\beq
\mathbb{P}_E[\tau_{S_{+}}<\tau_{S_{-}}]=\frac{1-N^{S_{-}}}{N^{S_{+}}-N^{S_{-}}}\lessapprox N^{-S_{+}},\label{PE}
\eeq
Thus Eq. \eqref{PE} is the probability that Eve will reach the value $S_+$ expected for Alice's authentication before reaching the value $S_-$ expected for Alice's non-authentication. To ensure that this probability is smaller than the specified tolerance $p_{fp}$, we may choose $S_{+}=S_{+}(N)$ to be 
\begin{equation}
S_{+}(N)=\left\lceil -\frac{\log p_{fp}}{\log N}\right\rceil,
\label{s+}
\end{equation}
where $\lceil x\rceil$ stands for the smallest integer greater than or equal $x\in\mathbb{R}$.

Similarly, the probability that Alice's $S$ parameter drifts to the low value $S_{-}$, and thus Alice is falsely not authenticated, is equal to 
\begin{equation}
\label{pfn}
\mathbb{P}_A[\tau_{S_{-}}<\tau_{S_{+}}]=\frac{1-\left(\frac{1-P_A}{P_A}\right)^{S_{+}}}{\left(\frac{1-P_A}{P_A}\right)^{S_{-}}-\left(\frac{1-P_A}{P_A}\right)^{S_{+}}}
\lessapprox \left(\frac{P_A}{1-P_A}\right)^{S_{-}}
\end{equation}
To ensure that this probability is smaller than the specified tolerance $p_{fn}$, we may choose 
\begin{equation}
S_{-}=S_{-}(N)=\left\lfloor\frac{\log p_{fn}}{\log \left(\frac{P_A}{1-P_A}\right)}\right\rfloor,
\label{s-}
\end{equation}
where $\lfloor x\rfloor$ stands for the greatest integer not exceeding $x\in\mathbb{R}$.
\section{Quantum advantage using quantum light}
We are now in position to explore the quantum advantage brought about by quantum light, in particular a single-photon source. Since the temporal summation window \cite{Holmes} of the visual system is on the order of $\tau_{\rm ts}\approx  400 {\rm ms}$, we can use a single-photon source, for example a heralded single-photon source \cite{herald1,herald2,herald3,herald4}, to produce a maximum of $\tilde{N}\sim 200$ photons within a time interval $\tau_{\rm ts}$, i.e. we need a production rate of at most 1 kHz, which is readily feasible. In the ideal case, the probability distribution of the number of photons incident on the cornea will be $p(n)=\delta_{n,\tilde{N}}$. 

It is known, however, that optical losses degrade the sub-Poissonian statistics of quantum light. Hence, if a pulse of light containing exactly $\tilde{N}$ photons propagates inside the lossy material of the eye, the statistics of the number of detected photons will not be a spiked distribution like $p(n)$, but a broader distribution. Since the optical losses quantified by the parameter $\alpha$ are not insignificant, the distribution of the detected photon number is not very far from the case of coherent light discussed previously. Yet, we will show that single photons do enhance performance, which demonstrates that quantum light sources can indeed be useful in addressing human vision. 

\begin{figure}
\begin{center}
\includegraphics[width=8. cm]{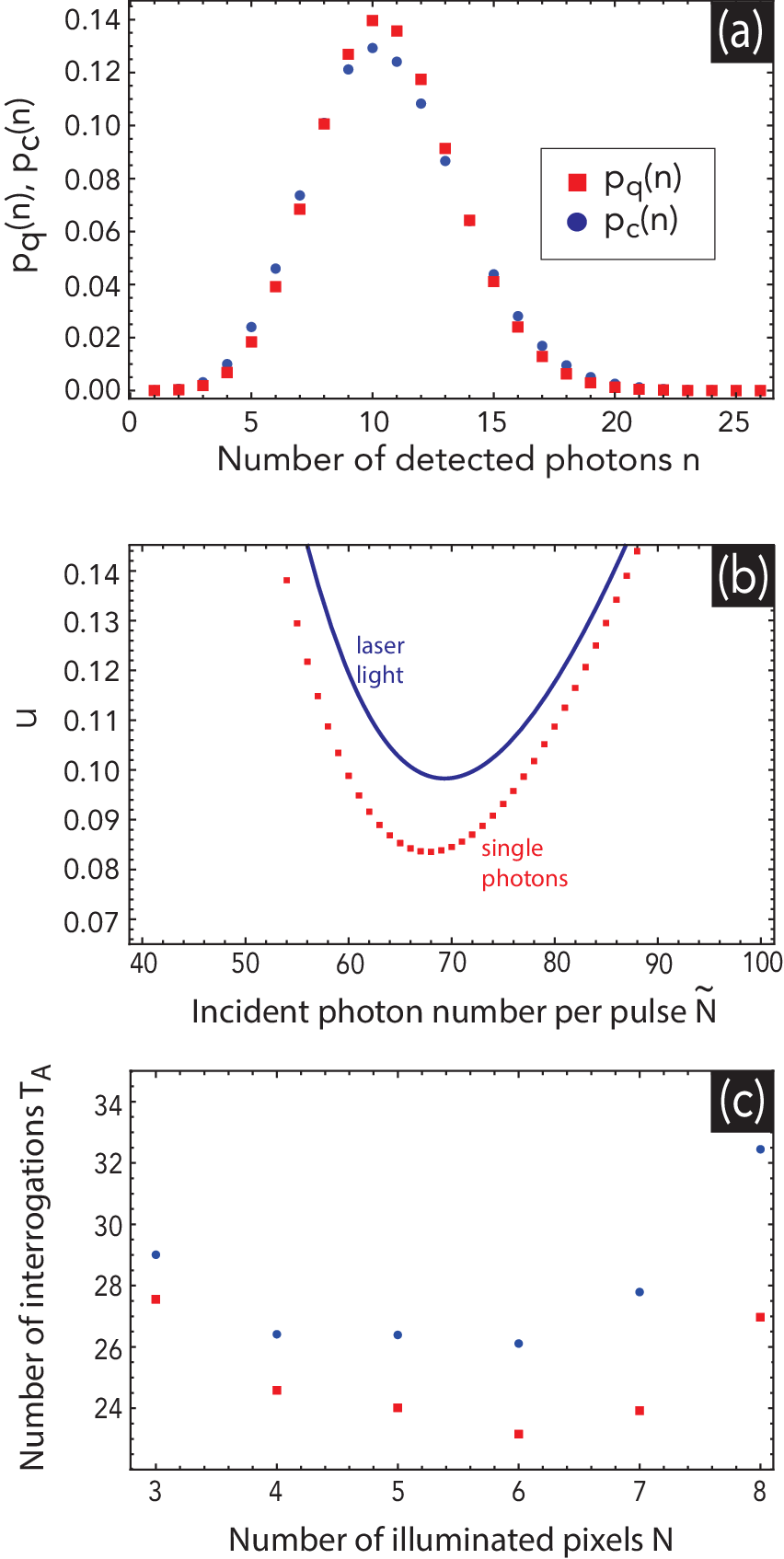}
\caption{Quantum advantage with single photons. (a) Example of the distribution of $n$, the photon number after transmission through the lossy material of loss parameter $\alpha$, which here is $\alpha=0.16$, and for an incident photon number $\tilde{N}=60$. For the incident light being coherent laser light, the distribution is $p_c(n)$, whereas for a pulse of 60 single photons incident on the eye the distribution is $p_q(n)$. The distribution $p_q(n)$ is slightly narrower than $p_c(n)$. (b) Example of the dependence of $u=p_H+p_L$ on the mean number of incident photons per pulse, $\tilde{N}$, for laser light and single photons. The minimum in each case determines the optimal mean photon number $\tilde{N}$. Here $\alpha_H=0.16$ and $\alpha_L=0.04$. (c) Interrogation time $T_A$ versus number of illuminated pixels $N$, for laser light and single photons. The minimum in each case determines the optimal number $N$ of pixels that should be illuminated, and the corresponding optimal interrogation time $T_A$. The quantum advantage is in this case a 11.4\% reduction in $T_A$. Blue circles refer to laser light and red squares to single photons.}
\label{T}
\end{center}
\end{figure}

In particular, for an optical loss parameter $\alpha$ and exactly $\tilde{N}$ photons incident on the cornea, the distribution of the photon number after propagation through the lossy material is binomial with parameters $(\tilde{N},\alpha)$, i.e.
\beq
p_q(n;\alpha)={{\tilde{N}!}\over {n!(\tilde{N}-n)!}}\alpha^n(1-\alpha)^{\tilde{N}-n},\qquad n=0,1,\ldots,\tilde{N},\label{pq}
\eeq
whereas, using Eq. 1, the corresponding distribution for coherent light is a Poissonian with parameter $\alpha\tilde{N}$, i.e. 
\begin{equation}
p_c(n;\alpha)=e^{-\alpha\tilde{N}}{({\alpha\tilde{N})^n}\over {n!}},\qquad n=0,1,\ldots\label{pc}
\end{equation}
where in this case $\tilde{N}$ is the mean incident number of photons per pulse. Note that in both cases the mean number of detected photons equals $\alpha\tilde{N}$. However, the variance of this quantity with coherent light is $V_c=\alpha\tilde{N}$, whereas the variance with quantum light is $V_q=\alpha (1-\alpha)\tilde{N}$. In Fig. \ref{T}a we show an example of the slight difference of the two distributions, Eqs. \eqref{pq} and \eqref{pc}, for $\alpha=0.16$ and $\tilde{N}=60$ photons.

Now, the probability that Alice fails to perceive an illuminated high-$\alpha$ spot is
\begin{equation}
p_H(\tilde{N})=\sum_{n=0}^{K-1}p(n;\alpha_H),
\end{equation}
whereas the probability that Alice does perceive an illuminated low-$\alpha$ spot is
\begin{equation}
p_L(\tilde{N})=1-\sum_{n=0}^{K-1}p(n;\alpha_L)
\end{equation}
In these expressions the probability mass function $p(n;\alpha)$ is equal to either $p_q(n;\alpha)$ of Equation \eqref{pq} for quantum light, or to $p_c(n;\alpha)$ of Equation \eqref{pc} for laser light. The parameter $u=p_H+p_L$ enters Eq. \eqref{pAp}, giving the probability that Alice succeeds in one particular interrogation. An example of the dependence of the parameter $u$ on the incident mean photon number $\tilde{N}$ is shown in Figure \ref{T}b. It is seen that $u$ is minimized at $\tilde{N}=69.4$ for laser light and $\tilde{N}=68$ for single photons, the minimum being $u_c=0.0983$ and $u_q=0.0839$, respectively (Note that for a coherent state relevant to laser light $\tilde{N}$ is a continuous variable). 
\subsection{Interrogation time}
Having found Alice's parameter $u$ for a given set of values $\alpha_H$ and $\alpha_L$, we can use Eq. \eqref{pAp} to determine the mean number, $T_A$, of interrogations required for Alice's authentication, and choose $N$ to minimize $T_A$. Indeed, by Theorem 1.3.5 in \cite{Norris} we find
\begin{equation}
\label{time}
T_A=\frac{S_{+}}{2P_A-1}-\frac{S_{+}-S_{-}}{2P_A-1}\mathbb{P}_A[\tau_{S_{-}}<\tau_{S_{+}}]\lessapprox\frac{S_{+}}{2P_A-1},
\end{equation}
where the second term in the exact expression for $T_A$ was neglected as it is proportional to $p_{fn}\ll 1$, and where $S_{+}=S_{+}(N)$ is given by \eqref{s+}, $S_{-}=S_{-}(N)$ is given by \eqref{s-}, and $P_A=P_A(N,u)$ is given by \eqref{pAp}. In Figure \ref{T}c we plot $T_A$ versus the number of illuminated retinal spots (pixels) $N$ for the two cases we consider, i.e. laser light and single-photons. It is seen that for laser light the optimal number of expected interrogations is $T_A\simeq 26$, with the optimal number of spots being $N=6$. For single-photons we find an optimal of approximately 23 expected interrogations with $N=6$ illuminated spots, i.e. we gain about 11.4\% in interrogation time. 

Finally, if we consider photon number as a resource, the fact that the optimal photon number per pulse for quantum light ($\tilde{N}=68$) is somewhat smaller than the corresponding number for laser light ($\tilde{N}=69.4$), pushes the total advantage to 13.3\%.
\section{Discussion}
We have here presented an intuitive biometric authentication process using the human visual system's ability to perform photon counting. The process rests on the perception of a number of illuminated spots on the retina. The performance of the authentication process in terms of false positive and false negative probability, as well as the interrogation time required to realize those two probabilities, is determined by the photon statistics of the stimulus light. We have shown that a single-photon source provides an advantage, even though the optical losses, which form the biometric fingerprint, degrade the sub-Poissonian statistics of the quantum stimulus light. Two comments are to be made. First, one might claim that the quantum advantage is small. However, a significantly larger quantum advantage cannot be excluded. This is because it is not straightforward to find the ultimate quantum limit of the general methodology, independent of the specific authentication strategy. Indeed, since the visual perception involves a number of nonlinearities, one might expect to be able to enhance the small difference in photon statistics between laser light and quantum light (at the photon detection level discussed previously) by designing another interrogation strategy. In the same context, one might argue that despite the quantum advantage shown here, the number of required interrogations is anyhow large. This is the reason we previously \cite{Loulakis} devised an authentication strategy based on pattern recognition. This strategy led to a much faster authentication time with the same performance metrics. However, this strategy depends on further assumptions on the workings of visual perception and pattern recognition. In contrast, with the simple strategy considered herein we aimed at a straightforward to understand and analytically tractable demonstration of the promise of quantum light sources for studying the human visual system in general, and our biometric authentication methodology in particular.
\appendix*
\section{}
The proof of the Lemma is as follows.
\begin{align}
\label{probab}
&\mathbb{P}[E_{fn}=E_{fp}]\nonumber\\
&={1\over {n+1}}\sum_{k=0}^n\mathbb{P}[E_{fn}=E_{fp} |\ H=k]\nonumber\\
&={1\over {n+1}}\sum_{k=0}^n\sum_{j=0}^k\mathbb{P}[E_{fn}=j | \ H=k]\ \mathbb{P}[E_{fp}=j |\ H=k]\nonumber\\
&={1\over {n+1}}\sum_{k=0}^n\sum_{j=0}^k{k\choose j}{n-k\choose j}p^j(1-p)^{k-j}q^j(1-q)^{n-k-j}
\end{align}
For a set $A\subset [n]$ of odd cardinality and median $k$, and for $x\in [n]$, we define
\beq
c(x,A)=\begin{cases} p,&\text{if } x\in A,\text{ and } x<k\\1-p,&\text{if } x\notin A,\text{ and } x<k\\1,&\text{if } x=k\\q,&\text{if } x\in A,\text{ and } x>k\\1-q,&\text{if } x\notin A,\text{ and } x>k\end{cases}
\eeq
and
$c(A)=\prod_{x=0}^n c(x,A)$. It is immediate that if $|A|=2j+1$, then $c(A)=p^j(1-p)^{k-j}q^j(1-q)^{n-k-j}$. On the other hand, there are precisely ${k\choose j}{n-k\choose j}$ sets $A\subset [n]$ with median $k$ and cardinality $|A|=2j+1$, since there are ${k\choose j}$ ways to choose the $j$ elements of $A$ in $\{0,\ldots,k-1\}$ and ${n-k\choose j}$ ways to choose the $j$ elements of $A$ in $\{k+1,\ldots,n\}$. Hence, with $\gamma_n:=\sum_{A\subset [n]\atop |A| \text{ odd}} c(A)$ it is 
\beq
\gamma_n=\sum_{k=0}^n\sum_{j=0}^k{k\choose j}{n-k\choose j}p^j(1-p)^{k-j}q^j(1-q)^{n-k-j}
\eeq
We can straightforwardly compute
\begin{equation}
\label{init}
\gamma_0=1\qquad \text{and}\qquad \gamma_1=1-p+1-q=2-(p+q).
\end{equation}
For $n\ge 2$, we have
\begin{align}
\gamma_n&=\sum_{\{0,n\}\subset A\subset[n]\atop {|A|\text{ odd}} }c(A)+\sum_{A\subset[n], \{0,n\}\cap A^c\neq\emptyset\atop {|A|\text{ odd}} }c(A)\nonumber\\
&=\sum_{\{0,n\}\subset A\subset[n]\atop {|A|\text{ odd}}}c(A)+\sum_{A\subset[n],\ 0\notin A\atop {|A|\text{ odd}} }c(A)\nonumber\\
&\label{recurse}+\sum_{A\subset[n],\ n\notin A\atop {|A|\text{ odd}} }c(A)-\sum_{A\subset[n],\ 0,n\notin A\atop {|A|\text{ odd}} }c(A)
\end{align}
If $0,n\in A$ then 
\[
c(A)=pq \prod_{x=1}^{n-1} c(x,A)=pq \prod_{x=0}^{n-2} c(x,\tilde{A})=pq\ c(\tilde{A}),\label{recur}
\]
where $\tilde{A}=A\setminus\{0,n\}-1 \subset [n\!-\!2]$. Hence,
\[
\sum_{\{0,n\}\subset A\subset[n]\atop {|A|\text{ odd}}}c(A)=pq \sum_{\tilde{A}\subset[n-2]\atop {|\tilde{A}|\text{ odd}}}c(\tilde{A})=pq\ \gamma_{n-2}.
\]
Likewise,
\begin{align}
\sum_{\{0,n\}\subset A^c\subset[n]\atop {|A|\text{ odd}}}c(A)&=(1-p)(1-q) \sum_{\tilde{A}\subset[n-2]\atop {|\tilde{A}|\text{ odd}}}c(\tilde{A})\nonumber\\
&=(1-p)(1-q) \gamma_{n-2},
\end{align}
and
\[
\sum_{A\subset[n],\ 0\notin A\atop {|A|\text{ odd}} }c(A)=(1-p)\ \gamma_{n-1},\qquad \sum_{A\subset[n],\ n\notin A\atop {|A|\text{ odd}} }c(A)=(1-q)\ \gamma_{n-1}.
\]
We may now substitute the preceding relations in \eqref{recurse} to get the recursive equation
\[
\gamma_n=(2-p-q)\gamma_{n-1}+(p+q-1)\gamma_{n-2},\qquad\text{for all }n\ge 2.
\]
In view of the initial condition \eqref{init}, the assertion of the Lemma can now be proved inductively, or simply by noting that $\gamma_n$ depends on $p,q$ only through $p+q$ and setting $q\leftarrow 0, \ p\leftarrow p+q$ in equation \eqref{probab}.\hfill$\Box$\\

\begin{acknowledgments}
This work has been co-financed by the European Union and Greek national funds through the Operational Program Competitiveness, Entrepreneurship and Innovation, under the call "RESEARCH-CREATE- INNOVATE", with project title "Photonic analysis of the retina's biometric photo-absorption" (project code: T1EDK-04921).
\end{acknowledgments}

\section*{Data Availability Statement}
Data available on request from the authors.

\end{document}